\documentclass[twocolumn,superscriptaddress,showpacs]{revtex4-1}
\usepackage{graphicx}
\usepackage{units}
\usepackage{multirow}
\usepackage{bigstrut}
\usepackage{overpic}
\usepackage{array}
\usepackage{color}
\usepackage{amsmath}
\usepackage{xspace}
\RequirePackage{lineno}
\newcommand{\sinTh}{\rm sin^2\theta_{2}/(1+cos^2\theta_{2})}
\newcommand{\avsinTh}{\rm \frac{\langle sin^2\theta_{2}\rangle }{\rm \langle 1+cos^2\theta_{2} \rangle }}
\newcommand{\dEdx}{\ensuremath{\mathrm{d}E/\mathrm{d}x}\xspace}
\newcommand{\grad}{\ensuremath{^{\circ}}}
\begin{document}

\title{\boldmath Measurement of Azimuthal Asymmetries in Inclusive Charged Dipion Production
in $e^+e^-$ Annihilations at $\sqrt{s}$ = 3.65\,GeV}

\author{
  \begin{small}
    \begin{center}
      M.~Ablikim$^{1}$, M.~N.~Achasov$^{9,f}$, X.~C.~Ai$^{1}$,
      O.~Albayrak$^{5}$, M.~Albrecht$^{4}$, D.~J.~Ambrose$^{44}$,
      A.~Amoroso$^{49A,49C}$, F.~F.~An$^{1}$, Q.~An$^{46,a}$,
      J.~Z.~Bai$^{1}$
% ...
      \\
      \vspace{0.2cm}
      (BESIII Collaboration)\\
      \vspace{0.2cm} {\it
        $^{1}$ Institute of High Energy Physics, Beijing 100049, People's Republic of China\\
% ...
$^{4}$ Bochum Ruhr-University, D-44780 Bochum, Germany\\
        $^{5}$ Carnegie Mellon University, Pittsburgh, Pennsylvania 15213, USA\\
% ...
$^{9}$ G.I. Budker Institute of Nuclear Physics SB RAS (BINP), Novosibirsk 630090, Russia\\
% ...
$^{44}$ University of Rochester, Rochester, New York 14627, USA\\
% ...
$^{46}$ University of Science and Technology of China, Hefei 230026, People's Republic of China\\
% ...
$^{49}$ (A)University of Turin, I-10125 Turin, Italy; (B)University of Eastern Piedmont, I-15121 Alessandria, Italy; (C)INFN, I-10125, Turin, Italy\\
% ...
$^{a}$ Also at State Key Laboratory of Particle Detection and Electronics, Beijing 100049, Hefei 230026, People's Republic of China\\
% ...
$^{f}$ Also at the Novosibirsk State University, Novosibirsk 630090, Russia\\
% ...
      }
    \end{center}
  \vspace{0.4cm}
  \end{small}
}


\begin{abstract}
We present a measurement of the azimuthal asymmetries of two charged pions in the inclusive process $e^+e^-\rightarrow \pi\pi X$,  based on a data set of 62\,$\rm{pb}^{-1}$ 
at the center-of-mass energy of $3.65$\,GeV collected with the BESIII detector.  
These asymmetries can be attributed to the Collins fragmentation function. 
We observe a nonzero asymmetry, which increases with increasing pion momentum.
As our energy scale is close to that of the existing semi-inclusive deep inelastic scattering experimental data, the measured asymmetries are important inputs for the global analysis of extracting the quark transversity distribution inside the nucleon and are valuable to explore the energy evolution of the spin-dependent fragmentation function.
\end{abstract}

\pacs{13.88.+e, 13.66.Bc, 13.87.Fh, 14.65.Bt}

\maketitle

The quark-hadron fragmentation process is parametrized with a fragmentation function (FF),
which describes the probability that a hadron carrying a fraction of the parton energy is found in the hadronization debris of the fragmenting parton.
The Collins FF, which considers the spin-dependent effects in fragmentation processes, was first discussed by Collins in Ref.~\cite{Collins}.
It connects the transverse quark spin with a measurable azimuthal asymmetry (the so-called Collins effect) in the distribution
of hadronic fragments along the initial quark's momentum.

The measurement of the Collins FF provides
an important test in understanding strong interaction dynamics and thus is of fundamental interest in 
understanding QCD, the underlying theory of the strong interaction. 
Because of its chiral-odd nature, 
it needs to couple to another chiral-odd function, for instance, the transversity distribution~\cite{Ralston,Jaffe-Ji1,Jaffe-Ji2} in semi-inclusive deep inelastic scattering (SIDIS) or another Collins FF in $e^+e^-$ annihilations,  
to form accessible observables. 
The transversity distribution, which contributes to the nucleon transverse spin, corresponds to the tensor charge of the nucleon 
and is the least known leading-twist quark distribution function.
There have been several SIDIS measurements of this asymmetry from HERMES~\cite{hermes1,hermes}, COMPASS~\cite{compass}, and JLab~\cite{Jlab}. 
Direct information on the  Collins FF can be obtained from $e^+e^-$ annihilation experiments, as suggested in Ref.~\cite{DBoer}. 
Measurements performed by the Belle~\cite{Belle1,Belle2,Belle3} and $BABAR$~\cite{BABAR} Collaborations give consistent nonzero asymmetries.
Based on the universality of the involved functions in $e^+e^-$ and SIDIS~\cite{Metz} experiments, global analyses~\cite{Anselmino,Martin}
have been performed to simultaneously extract the transversity and Collins FF.
However, the  $e^+e^-$ Collins asymmetries taken from Belle and $BABAR$ correspond to considerably higher $Q^2$ ($\approx
100\,\text{GeV}^2$) than the typical energy scale of the existing SIDIS data (mostly 2$-$20\,$\text{GeV}^2$). 
Therefore, the energy evolution of the Collins FF at different $Q^2$ is a key factor to evaluate the transversity~\cite{Kang}.
Recently, the treatment of the evolution is developed in Refs~\cite{Peng,Peng2,Feng,Kang2,Kanazawa}, which predict about a factor of 2 change in the
observed asymmetries between BESIII energy and Belle and $BABAR$ energy, but is not directly validated by experimental data.
The BESIII experiment~\cite{BESIII} studies $e^+e^-$ annihilations at a moderate energy scale (4$-$20\,$\rm GeV^2$).
It is important to investigate the interesting feature of Collins FF at this energy scale,  and the results can then be connected more directly to the SIDIS.
Moreover, as emphasized in Ref.~\cite{Peng}, with significantly lower $Q^2$ with respect to $B$ factories,  the results will be crucial to explore the $Q^2$ evolution of the Collins FF and further the uncertainty of the extracted transversity, thus improving our understanding of both Collins FF and transversity. 

In this Letter, we present the measurement of azimuthal
asymmetries in hadron-hadron correlations for inclusive charged pion pair production  $e^+e^-\rightarrow\pi\pi X$,
which can be attributed to the Collins effect.
The analysis is based on a data sample with
an integrated luminosity of 62\,pb$^{-1}$ collected with
the BESIII detector~\cite{BESIII} at the center-of-mass energy
$\sqrt{s}=3.65$\,GeV, where the energy is away from resonances. 
Compared to the existing $e^+e^-$ data, in this measurement, only fragmenting $u, d, s$ quarks are involved.
The results are free from charm contribution; as such, the combination with SIDIS data is more
straightforward. 
The apparatus relevant to this work includes a main drift chamber (MDC), a time-of-flight system, and an electromagnetic calorimeter (EMC). Details on the features and capabilities of the BESIII detector can be found in Refs.~\cite{BEPCII,BESIII}.

Monte Carlo (MC) simulated events, which are processed with a full
{\sc geant}4-based~\cite{Geant4} simulation of the BESIII detector,
are used to optimize the event selection criteria and check for systematics.
The MC samples for light quarks in $e^+e^-\rightarrow q\bar{q}$ $(q=u,d,s)$ processes are generated by the {\sc luarlw}~\cite{luarlw} package,
which is based on the Lund model~\cite{Lund,Lund2}.
More MC samples including QED processes ($e^+e^-\rightarrow l^+l^-$ ($l=e,\mu,\tau$), $e^+e^-\rightarrow\gamma\gamma$),
two photon fusion ($e^+e^-\rightarrow e^+e^- X$), line-shape tail production of $\psi(2S)$ and, and the initial state radiative
process $e^+e^-\rightarrow \gamma J/\psi$ are analyzed to identify possible backgrounds.

Taking into account the spin of the quark, the number density $D_{h}^{q\uparrow}$ for finding a spinless hadron $h$ with
transverse momentum $\textbf{P}^{\perp}_{h}$ produced from a transversely polarized quark $q$
with spin ${\bf{S}}_{q}$ can be described in terms of the unpolarized FF $D_{1}^{q}$ and
the Collins FF $H_{1}^{\perp q}$ at the leading twist~\cite{Bacchetta}
\begin{equation}
D_{h}^{q\uparrow}(z,{\bf{P}}^{\perp}_{h})=D_{1}^{q}(z,{\bf{P}}^{\perp 2}_{h}) + H_{1}^{\perp q}(z,{\bf{P}}^{\perp 2}_{h})\frac{(\hat{\bf{k}} \times {\bf{P}}^{\perp}_{h}) \cdot {\bf{S}}_{q} } { z M_{h} },
\label{eq-1}
\end{equation}
where $\hat{\bf{k}}$ denotes the direction of the initial quark $q$, $z={\rm{2}}E_h/Q$ denotes
the fractional energy of the hadron relative to half of $Q=\sqrt{s}$,
and $M_{h}$ is the hadron mass.
The second term contains the Collins FF and depends on the spin orientation of the
quark $q$, which leads to a sine modulation of the angle spanned by $\textbf{P}^{\perp}_{h}$ and the plane normal to the quark spin.

\begin{figure}
\includegraphics[width=0.28\textwidth]{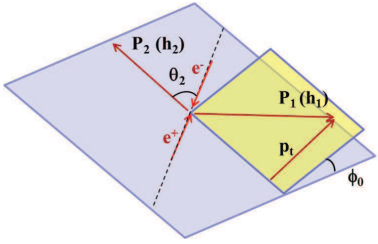}
\caption{The angle $\phi_{0}$ is defined as the angle between the plane spanned by the beam axis and the momentum of the second hadron ($P_{2}$)
and the plane spanned by the transverse momentum $p_{t}$ of the first hadron relative to the second hadron.
The angle $\theta_2$ is the polar angle of the second hadron.
\label{phi0_define}}
\end{figure}

In hadron production in $e^+e^- \to q \bar{q}$  events, the Collins effect can be observed when the fragments of the quark and antiquark are considered simultaneously. 
At $\sqrt{s}=3.65$\,GeV,  due to the absent of the clear jet structure,  there is no good way to estimate the $q$-$\bar{q}$ axis.
However, the Collins asymmetries can be investigated 
with the azimuthal angle $\phi_0$ defined as
the angle between the plane spanned by the beam axis and the momentum of the second hadron ($P_{2}$),
and the plane spanned by the transverse momentum $p_{t}$ of the first hadron relative to the second hadron~\cite{DBoer,DBoer2}, as shown in Fig.~\ref{phi0_define}.

The normalized dihadron yield is recorded as a function of $\phi_0$ and can be parametrized as $a\cos{(2\phi_{0})}+b$, with $b$
referring to the term which is independent of $\phi_{0}$, and $a$ can be written as~\cite{DBoer,DBoer2}
\begin{equation}
a(\theta_2,z_1,z_2) = \frac{\sin^2\theta_2}{1+\cos^2\theta_2} \frac{\mathcal{F}\left( H_1^{\perp}(z_1)\bar{H}_1^{\perp}(z_2)/M_1M_2\right)}{\mathcal{F}\left(D_1(z_1)\bar{D}_1(z_2)\right)},
\label{eq-2}
\end{equation}
where $\mathcal{F}$ denotes a convolution over the $p_{t}$. The $M_1$ and $M_2$
are the masses of the two hadrons, $z_1$ and $z_2$ are their fractional energies, and $\theta_2$ is the polar angle of the second hadron with respect to the beam axis.
$\bar{D}_1$ and $\bar{H}_1^{\perp}$ denote FFs for antiquarks.

We reconstruct charged tracks from hits in the MDC. We require the polar angle in the laboratory frame to satisfy $|\cos\theta|<0.93$,
and the point of closest approach to the interaction vertex of $e^+e^-$ is required to be within 1\,cm
in the plane transverse to the beam line and within 10\,cm along the beam axis. 
Particle identification for charged tracks is accomplished by combining the measured energy loss
(\dEdx) in the MDC and the flight time obtained from the time-of-flignt to determine a probability $\mathcal{L}$($h=K, \pi$, $p$, $e$)
for each particle ($h$) hypothesis.
The $\pi^\pm$($K^\pm$) candidates are required to satisfy $\mathcal{L}(\pi)(\mathcal{L}(K))>0.001$, $\mathcal{L}(\pi)>\mathcal{L}(K)$ ($\mathcal{L}(K)>\mathcal{L}(
\pi)$)and $\mathcal{L}(\pi)(\mathcal{L}(K))>\mathcal{L}(p)$.
Electrons are identified with the requirement $\mathcal{L}(e)>0.001$ and the ratio  $\mathcal{L}(e)/(\mathcal{L}(e)+\mathcal{L}(\pi) + \mathcal{L}(K))>0.8$.
Photons are reconstructed from isolated clusters in the EMC, whose energies are required to be larger than 25\,MeV
in the EMC barrel region ($|\cos\theta|< 0.8$) and 50\,MeV in end caps ($0.84 <|\cos\theta|< 0.92$).
It is required that the cluster timing delay from the reconstructed event start time does not exceed 700\,ns in order to suppress
electronic noise and energy deposits unrelated to the event.
To select inclusive $e^+e^-\rightarrow\pi\pi X$ events,
at least three charged tracks are required in order to strongly suppress two body decays.
At least two of the charged tracks should be identified as pions.
To suppress QED backgrounds with the final state $\tau^+\tau^-$ and unphysical backgrounds, e.g. beam-gas interactions,
the visible energy in the detector, which is defined as the total
energy of all reconstructed charged tracks and photons, is required to be larger than 1.5\,GeV and
no electron must be present in the event.
Studies based on MC samples indicate that the backgrounds are suppressed to a negligible level, less than 2.5\%.
We select pion pairs with $z_{1(2)} \in [0.2, 0.9]$,
where the lower bound is used to reduce pions originated from resonance decays (mostly $\rho$, $f$), and the upper bound is used to reject
two body decays.
Compared to measurement at a higher energy scale~\cite{Belle1,BABAR}, there is no clear jet event shape at BESIII
which could help to separate the hadrons coming from different fragmenting\linebreak[4] (anti)quark. Instead, to select back-to-back pions, we require the opening angle of the two charged pion candidates to be larger than $120{\grad}$.
This requirement reduces the possibility that two pions come from the fragmentation of the same quark.
We label the two pions randomly as $h_{1}$ and $h_{2}$, and we use the momentum direction of $h_{2}$ as reference axis.
If more than two pions are present in an event, they are combined to each other, which means each pion is allowed to be assigned to different pion pairs.
In the final event selection, 331\,696 events survived, which provide 557\,204 available charged pion pairs.

We introduce the $2\phi_{0}$ normalized ratio, $R = \frac{N(2\phi_0)}{\langle N_0 \rangle}$,
where $N(2\phi_0)$ is the dipion yield in each (2$\phi_0$) subdivision, and $\langle N_0 \rangle$ is the averaged bin content.
The normalized ratios are built for unlike sign ($\pi^{\pm}\pi^{\mp}$), like sign ($\pi^{\pm}\pi^{\pm}$) and
all pion pairs ($\pi\pi$), defined as $R^{\rm U}$, $R^{\rm L}$ and $R^{\rm C}$, respectively,
in which different combinations of favored FFs and disfavored FFs are involved.
A favored fragmentation process refers to the fragmentation of a quark into a hadron containing a valence quark
of the same flavor, for example, $u(\bar{d})\rightarrow\pi^+$,
while the corresponding $u(\bar{d})\rightarrow\pi^-$ is a disfavored process.
Since the normalized ratio $R$ is strongly affected by detector acceptance,
we use double ratios $R^{\rm U}/R^{\rm L(C)}$ (UL and UC ratios)~\cite{Belle1,Belle2}
to extract the azimuthal asymmetries. 
The gluon radiation may induce a $\cos(2\phi_0)$ modulation according to Ref.~\cite{DBoer2}, but
it is highly suppressed at the BESIII energy scale and is independent of the charge of the pions.
Through the double ratios, charge-independent instrumental effects cancel out, and QCD radiative effects are negligible at the first order,
while the charge-dependent Collins asymmetries are kept.
The double ratio $R^{\rm U}/R^{\rm L(C)}$ follows the expression
\begin{equation}
\frac{R^{\rm U}}{R^{\rm L(C)}} = A \cos(2\phi_{0}) + B,
\label{eq-3}
\end{equation}
where $A$ and $B$ are free parameters. $B$ should be consistent with unity,
and $A$ mainly contains the Collins effect.
The $A_{\rm UL}$, $A_{\rm UC}$ are used to denote the asymmetries for UL and UC ratios, respectively.

\begin{figure}
\includegraphics[width=0.43\textwidth]{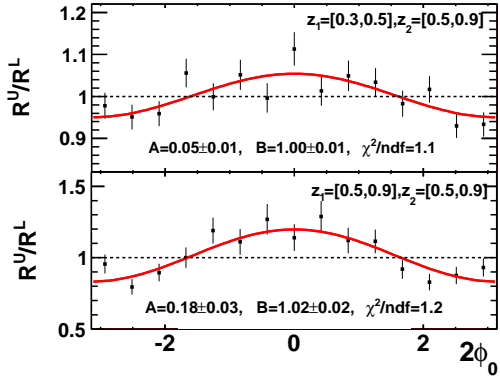}
\caption{Double ratio $R^{\rm U}/R^{\rm L}$ versus  2$\phi_{0}$ in the bin $z_{1}\in[0.3, 0.5]$, $z_2\in$[0.5, 0.9] (top) and bin $z_{1}\in[0.5, 0.9]$, $z_2 \in[0.5, 0.9]$ (bottom). The solid lines show the results of the fit.\label{fit_show}}
\end{figure}

\begin{figure}
\includegraphics[width=0.45\textwidth]{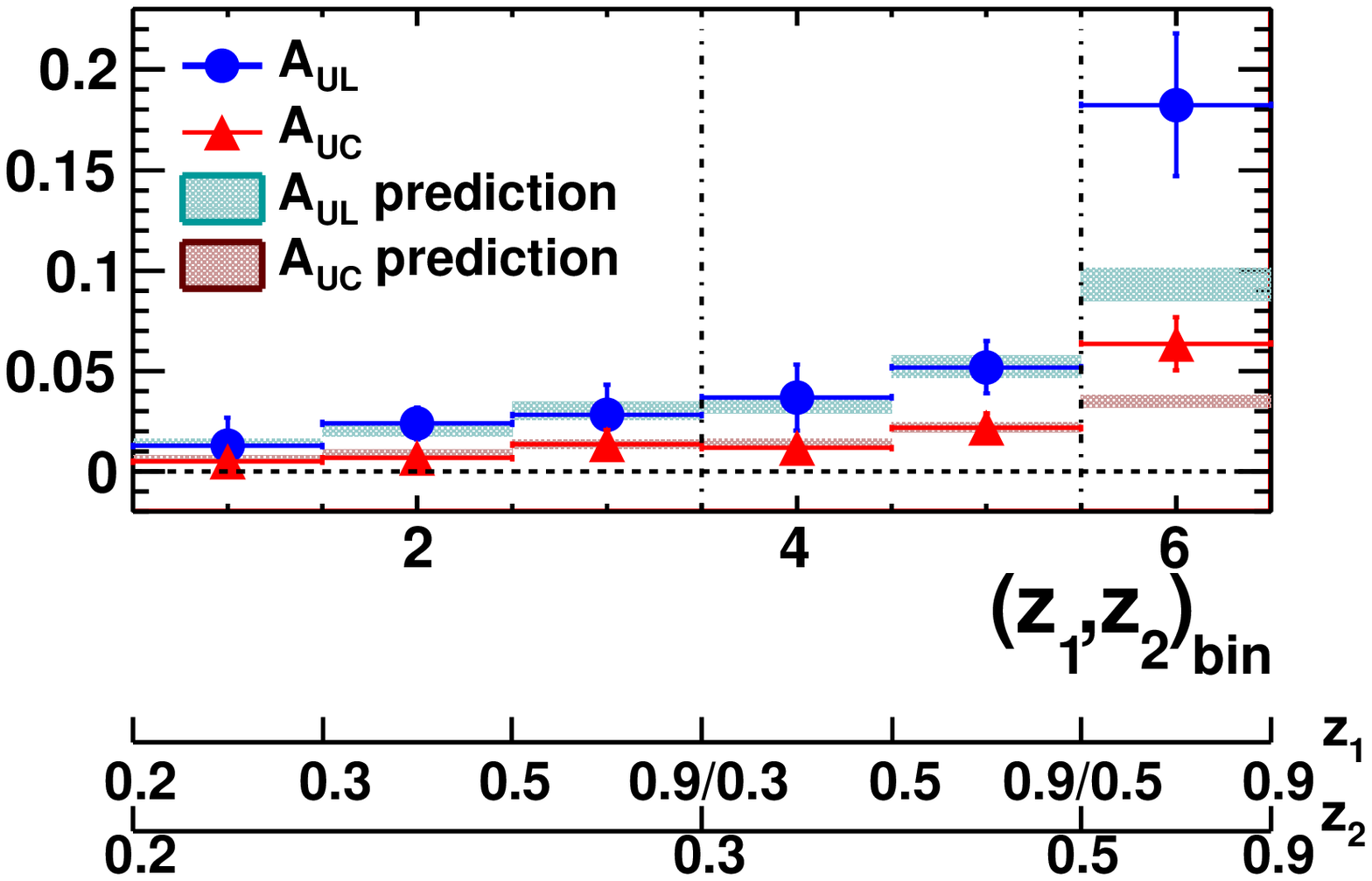}
\includegraphics[width=0.45\textwidth]{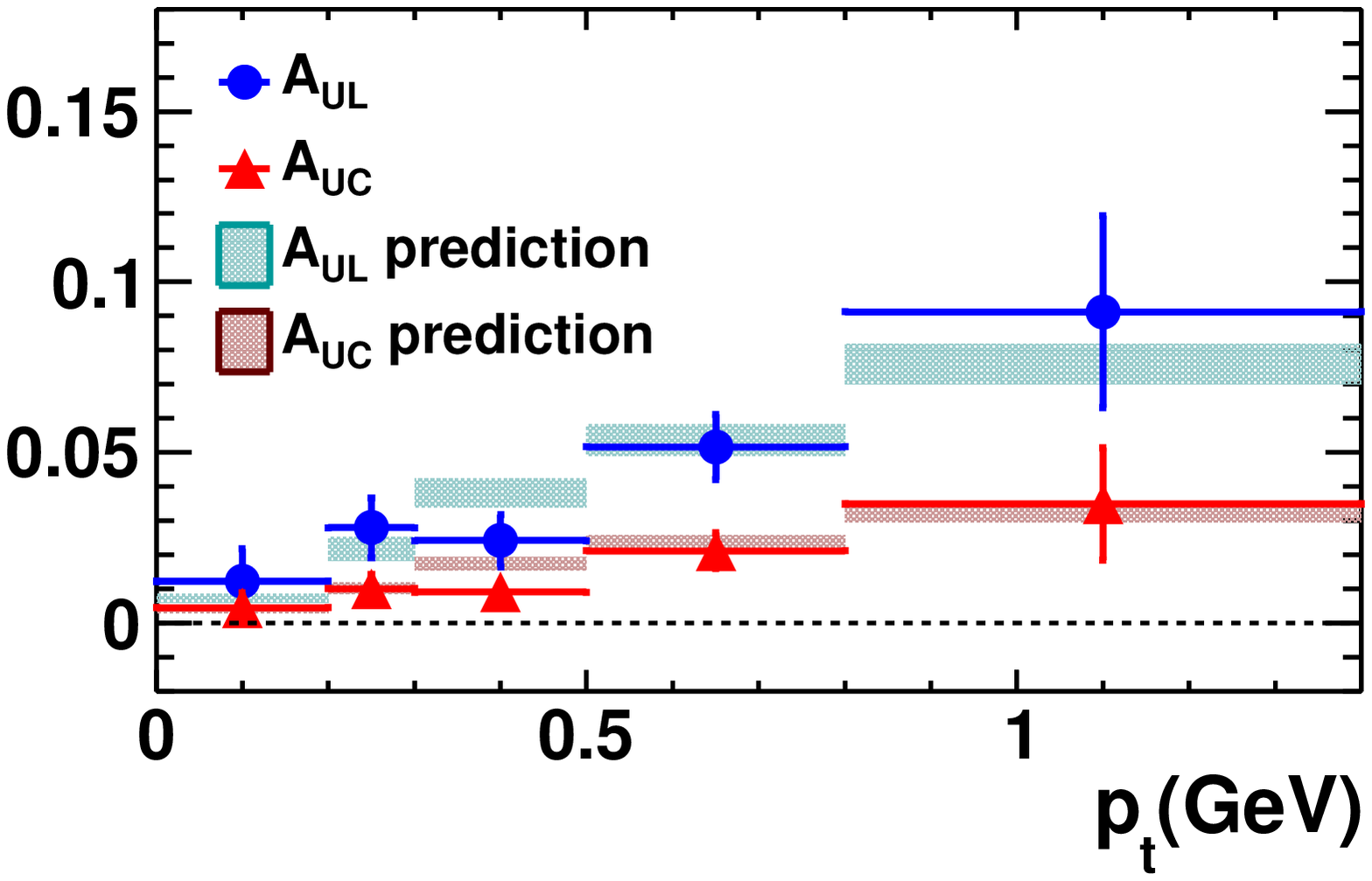}
\caption{Asymmetries as a function of fractional energies ($z_1,z_2$) (top) and $p_{t}$ (bottom) for the UL (dots) and UC (triangles) ratios, where the $p_{t}$ refers to the transverse
 momentum of the first hadron relative to the second hadron, as shown in Fig.~\ref{phi0_define}.
In the top figure, the lower scales show the boundaries of the bins in $z_1$ and $z_2$. Theoretical predictions from the authors of  Ref.~\cite{Peng2} are overlaid, 
where the hatched areas show the predicted bands.
\label{fig:vsZbin}
\label{fig:vsPt}}
\end{figure}

\begin{figure}
\includegraphics[width=0.43\textwidth]{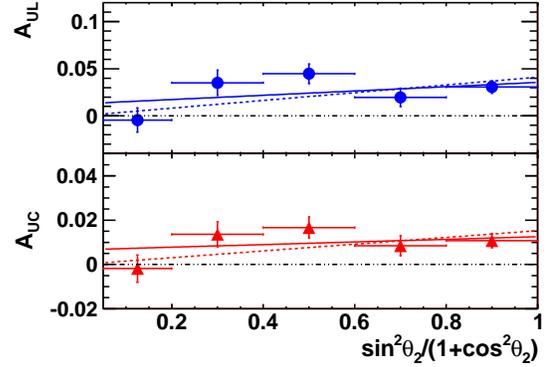}
\caption{Asymmetries as a function of $\sinTh$ for UL (dots) and UC (triangles) ratios. 
Linear fits with the constant term being set to 0 (dashed line) or a free parameter (solid line) are shown.
\label{fig:vsSin}}
\end{figure}

The analysis is performed in bins of ($z_1$, $z_2$), $p_t$ and $\sinTh$.
In ($z_1$, $z_2$) bins, the boundaries are set at $z_{i}$= 0.2, 0.3, 0.5 and 0.9 ($i=1,2$), where complementary off-diagonal bins ($z_1$, $z_2$) and ($z_2$, $z_1$) are combined.
In each bin, normalized rates $R^{\rm U,L,C}$ and double ratios $R^{\rm U}/R^{\rm L,C}$ are evaluated.
In Fig.~\ref{fit_show}, the distributions of the double ratio $R^{\rm U}/R^{\rm L}$ are shown as an example for two highest ($z_1$, $z_2$) bins with the fit results using Eq.~(\ref{eq-3}).
The asymmetry values ($A$) obtained from the fits are shown as a function of six symmetric ($z_1$, $z_2$) bins, $p_t$ and $\sinTh$ bins in Figs.~\ref{fig:vsZbin} and \ref{fig:vsSin}, respectively.
The numerical results in each ($z_{1}$,$z_{2}$) and $p_t$ bins are listed in Table~\ref{table:results_raw}.



Several potential sources of systematic uncertainties are investigated and all systematic uncertainties are added in quadrature finally.
An important test is the extraction of
double ratios from MC samples, in which the Collins asymmetries
are not included but radiative gluon and detector acceptance effects are taken into account.
In the MC samples, which is about 10 times of data statistics, double ratios are found to be consistent with 0 in all bins within statistical uncertainties.
To test any potential smearing effects in the reconstruction process,
MC samples are reweighted to produce generated asymmetries which vary in (0.02, 0.15) for
UL ratios and (0.01, 0.08) for UC ratios in different bins.
The reconstructed asymmetries are basically consistent with input, the differences between them,
which range from 0.2\% to 48\% for UL ratios and range from 2\% to 31\% for UC ratios relatively,
are included in the systematic uncertainties.

Additional possible contribution from gluon radiation can be examined in data by subtracting the normalized
yields, $R^{\rm U}-R^{\rm L(C)}$.
The subtraction method will cancel all the radiative terms, but the cancellation of the acceptance effects may be incomplete.
The differences between the asymmetries obtained with the subtraction method and the nominal results
range from 0.001 to 0.01 for UL ratios and from 0.0 to 0.005 for UC ratios.  These are assigned as absolute systematic uncertainties.

The probability of misidentifying kaons as pions may introduce $K\pi$ pairs and $KK$ pairs into the $\pi\pi$ samples of interest.
However, due to the much lower inclusive production cross section for charged kaons compared to pions,
the $\pi$$\pi$ asymmetry receives non-negligible contribution only from the $K$$\pi$ combination.
We denote with $A^{\pi\pi}$ and $A^{K\pi}$ the corresponding Collins asymmetries in data.
They can be obtained by unfolding the measurements of $A^{\pi\pi}_{\rm mea.}$ and $A^{K\pi}_{\rm mea.}$, where
$A^{\pi\pi}_{\rm mea.} = ( 1 - f_{K\pi}) A^{\pi\pi} + f_{K\pi} A^{K\pi},$
$A^{K\pi}_{\rm mea.} = ( 1 - f_{\pi\pi}) A^{K\pi} + f_{\pi\pi} A^{\pi\pi},$
$f_{K\pi}$ and $f_{\pi\pi}$ are the MC-determined contamination fractions.
Depending on the $(z_1, z_2)$ bin, $f_{K\pi}$ is found to range from 0.0\% to 4.5\% and $f_{\pi\pi}$ ranges from 0.1\% to 35.4\%,
The errors on $A^{K\pi}$ are very large, and the changes in $A^{\pi\pi}$ from the nominal values are in (0.001, 0.005) for UL ratios and (0.0, 0.001) for UC ratios, and are assigned as systematic uncertainty. 

Additional higher harmonic terms (such as $\sin\rm{2} \phi_0$ and $\cos\rm{4} \phi_0$)
are also included in the fit function to validate the robustness of the fit.
The changes of the value of the cosine asymmetries, which vary in (0.001, 0.009) for UL ratios and
(0.0, 0.003) for UC ratios, are included in the systematic uncertainties.

We have also verified null asymmetries for the double ratio of $\pi^+\pi^+$/$\pi^-\pi^-$ pairs and for random
combinations of pairs of tracks from different events.
From these tests, no significant asymmetries are observed.
The beam polarization may contribute to the measured asymmetries.
We study the angular distribution of the $e^{+}e^{-}\rightarrow \mu^{+}\mu^{-}$ process, which is sensitive to beam polarization.
No buildup of polarization is observed.


Adding statistical and systematic uncertainties in quadrature, we observe significant, nonzero Collins
asymmetries, as shown in Fig.~\ref{fig:vsZbin}.
These asymmetries rise with fractional energies and $p_t$ as expected theoretically~\cite{DBoer} and seen in higher-energy $e^+e^-$
experiments~\cite{Belle1,Belle2,Belle3,BABAR}. 
The predictions of authors of Ref.~\cite{Peng2}, based on results from previous data and the energy evolution model,
are also shown in  Fig.~\ref{fig:vsZbin}, and are basically consistent with our results. 
A direct comparison with higher-energy $e^+e^-$ data is meaningless due to differing kinematics. 
However, asymmetries in our data are $1.5$ times higher overall and higher by 0$-$2 sigma at points of comparable $z$ and $p_t$.

The expected behavior of the Collins asymmetries as a function of $\sinTh$ is linear and vanishes at $\theta_2 = 0$, as formulated in Eq.~(\ref{eq-2}).
Thus, a linear fit is performed to the points in Fig.~\ref{fig:vsSin},
with the constant term set to be 0 or left as a free parameter,
which gives the reduced $\chi^{2}$ to be 2.3 or 2.8 for $A_{\rm UL}$ and 1.7 or 1.9 for $A_{\rm UC}$
respectively. 
The significance for a zero offset is only about 1$\sigma$ for both $A_{\rm UL}$ and $A_{\rm UC}$.

The authors of the very recent paper Ref.~\cite{Peng2} give the theoretical predictions for the BESIII energy scale, which are also shown in Fig.~\ref{fig:vsZbin}.
Overall, our measured asymmetries are compatible with those predictions, except at the largest $z$ interval.

\begin{widetext}

\begin{table} 
\caption{Results of $A_{\rm UL}$ and $A_{\rm UC}$ in each ($z_{1},z_{2}$) and $p_{t}$ bin. The uncertainties are statistical and systematic, respectively. The averages $\langle z_i\rangle$, $\langle p_t\rangle$ and $\avsinTh$ are also given.\label{table:results_raw}}
\begin{ruledtabular}
\begin{tabular}{ccccccc}
 $z_1$ $\leftrightarrow$ $z_2$ & $\langle z_1\rangle$ & $\langle z_2\rangle$ & $\langle p_t\rangle$\,(GeV) & $\frac{\langle \sin^2\theta_2\rangle}{\langle 1+\cos^2\theta_2 \rangle}$ &  $A_{\rm UL}$ & $A_{\rm UC}$\\
\hline
$[0.2, 0.3][0.2, 0.3]$   & 0.245 & 0.245 & 0.262 & 0.589 & 0.0128  $\pm$ 0.0085 $\pm$ 0.0114  & 0.0050 $\pm$ 0.0038  $\pm$ 0.0017   \\
$[0.2, 0.3][0.3, 0.5]$   & 0.311 & 0.311 & 0.329 & 0.576 &  0.0240  $\pm$ 0.0068 $\pm$ 0.0042  & 0.0067 $\pm$ 0.0032 $\pm$ 0.0041 \\
$[0.2, 0.3][0.5, 0.9]$   & 0.428 & 0.426 & 0.444 & 0.572 & 0.0281  $\pm$ 0.0131 $\pm$ 0.0077  & 0.0136 $\pm$ 0.0064  $\pm$ 0.0029   \\
$[0.3, 0.5][0.3, 0.5]$   & 0.379 & 0.379 & 0.388 & 0.563 & 0.0369  $\pm$ 0.0097 $\pm$ 0.0132  & 0.0117 $\pm$ 0.0046  $\pm$ 0.0015   \\
$[0.3, 0.5][0.5, 0.9]$   & 0.498 & 0.499 & 0.479 & 0.564 & 0.0518  $\pm$ 0.0120 $\pm$ 0.0049  & 0.0217 $\pm$ 0.0056  $\pm$ 0.0046   \\
$[0.5, 0.9][0.5, 0.9]$   & 0.625 & 0.628 & 0.499 & 0.570 & 0.1824  $\pm$ 0.0290 $\pm$ 0.0204  & 0.0637 $\pm$ 0.0118  $\pm$ 0.0061   \\
\hline
\hline
$p_{t}$\,(GeV) & $\langle p_{t} \rangle $\,(GeV) & $\langle z_1 \rangle $ & $\langle z_2 \rangle $ & $\avsinTh$ & $A_{\rm UL}$ & $A_{\rm UC}$ \\
\hline
$[0.00, 0.20]$ & 0.133 & 0.291 &  0.348 & 0.574 &  0.0122 $\pm$ 0.0093 $\pm$ 0.0021 &  0.0044 $\pm$ 0.0043  $\pm$ 0.0006 \\
$[0.20, 0.30]$ & 0.253 & 0.285 &  0.344 & 0.579 &  0.0279 $\pm$ 0.0081 $\pm$ 0.0034 &  0.0100 $\pm$ 0.0038  $\pm$ 0.0016 \\
$[0.30, 0.45]$ & 0.405 & 0.327 &  0.346 & 0.570 &  0.0241 $\pm$ 0.0072 $\pm$ 0.0025 &  0.0090 $\pm$ 0.0031  $\pm$ 0.0026 \\
$[0.45, 0.80]$ & 0.610 & 0.453 &  0.349 & 0.571 &  0.0516 $\pm$ 0.0087 $\pm$ 0.0040 &  0.0211 $\pm$ 0.0049  $\pm$ 0.0019 \\
$[0.80, 1.40]$ & 0.923 & 0.646 &  0.334 & 0.584 &  0.0913 $\pm$ 0.0249 $\pm$ 0.0133 &  0.0350 $\pm$ 0.0116  $\pm$ 0.0116 \\
\end{tabular}
\end{ruledtabular}
\end{table}

\end{widetext}

In summary, we perform a measurement of the azimuthal asymmetry 
in the inclusive production of charged pion pairs. 
Our results suggest nonzero asymmetry in the region of large fractional energy $z$,  which can be attributed to the product of a quark and an antiquark Collins function. 
This is the first measurement of the Collins asymmetry at
low energy scale ($Q^2 \approx 13\,\rm GeV^2$) in  $e^+e^-$ annihilation.    
The observed  asymmetry indicates a larger spin-dependent Collins effect than those at the higher energy scale from $B$ factories~\cite{Belle1,Belle2,Belle3,BABAR}.
The results are of great importance to explore the $Q^2$ evolution of the Collins function and extract transversity distributions in nucleon.

\begin{acknowledgments}
The authors would like to thank D. Boer, X. D. Jiang, J. P. Ma, P. Sun and F. Yuan for helpful discussions on the theoretical aspects of the measurement.
The BESIII Collaboration thanks the staff of BEPCII and the IHEP computing center for their strong support. This work is supported in part by National Key Basic Research Program of China under Contract No. 2015CB856700; National Natural Science Foundation of China (NSFC) under Contracts Nos. 11125525, 11235011, 11275266, 11322544, 11335008, 11425524; the Chinese Academy of Sciences (CAS) Large-Scale Scientific Facility Program; the CAS Center for Excellence in Particle Physics (CCEPP); the Collaborative Innovation Center for Particles and Interactions (CICPI); Joint Large-Scale Scientific Facility Funds of the NSFC and CAS under Contracts Nos. 11179007, U1232201, U1332201; CAS under Contracts Nos. KJCX2-YW-N29, KJCX2-YW-N45; 100 Talents Program of CAS; National 1000 Talents Program of China; INPAC and Shanghai Key Laboratory for Particle Physics and Cosmology; German Research Foundation DFG under Contract No. Collaborative Research Center CRC-1044; Istituto Nazionale di Fisica Nucleare, Italy; Ministry of Development of Turkey under Contract No. DPT2006K-120470; Russian Foundation for Basic Research under Contract No. 14-07-91152; The Swedish Resarch Council; U. S. Department of Energy under Contracts Nos. DE-FG02-04ER41291, DE-FG02-05ER41374, DE-FG02-94ER40823, DESC0010118; U.S. National Science Foundation; University of Groningen (RuG) and the Helmholtzzentrum fuer Schwerionenforschung GmbH (GSI), Darmstadt; WCU Program of National Research Foundation of Korea under Contract No. R32-2008-000-10155-0.
\end{acknowledgments}

\bibliography{mybib}

\end{document}